\documentclass[aps,pra,twocolumn,showpacs,groupedaddress]{revtex4}

\usepackage{graphicx}
\usepackage{amsmath}
\usepackage{amssymb}
\usepackage{bm}
\usepackage{color}
\usepackage{array}
\usepackage{pdfpages}

\begin{document}
\title{Doppler Spectroscopy of an Ytterbium Bose-Einstein Condensate on the Clock Transition}
\author{A. Dareau}
\author{M. Scholl}
\author{Q. Beaufils}
\author{D. D\"oring}
\altaffiliation{Current address: FARO Scanner Production GmbH, Lingwiesenstrasse 11/2, D-70825 Korntal-M{\"u}nchingen}
\author{J. Beugnon}
\author{F. Gerbier} 
\affiliation{Laboratoire Kastler Brossel, Coll\`ege de France, CNRS, ENS-PSL Research
University, UPMC-Sorbonne Universit\'es, 11 place Marcelin-Berthelot, 75005 Paris}
\date{\today}
\begin{abstract}
We describe Doppler spectroscopy of Bose-Einstein condensates of ytterbium atoms using a narrow optical transition. We address the optical clock transition around 578 nm between the ${^1}S_0$ and  ${^3}P_0$  states with a laser system locked on a high-finesse cavity. We show how the absolute frequency of the cavity modes can be determined within a few tens of kHz using high-resolution spectroscopy on molecular iodine. We show that optical spectra reflect the velocity distribution of expanding condensates in free fall or after releasing them inside an optical waveguide. We demonstrate sub-kHz spectral linewidths, with long-term drifts of the resonance frequency well below 1\,kHz/hour. These results open the way to high-resolution spectroscopy of many-body systems.
\end{abstract}
\maketitle
%
%
\section{Introduction}
Since its early days, the field of ultracold atoms has traditionally been focused on alkali atoms, which feature a strong, closed transition favorable for laser cooling and trapping. In the last ten years, a lot of progress has been made in bringing more complex atoms to quantum degeneracy \cite{takasu2003a,fukuhara2007b,escobar2009a,stellmer2009a,kraft2009a}. Two-electron atoms, for instance from the alkaline-earth family or ytterbium, have been of particular interest because of one special feature, an ultranarrow $J=0 \rightarrow J'=0$ optical transition from the ${^1}S_0$ ground state to a metastable ${^3}P_0$ optically excited state  (here $J,J'$ denote the electronic angular momentum for the ground and excited states). Although electric dipole-forbidden, the transition can be weakly enabled by hyperfine interactions or an applied magnetic field \cite{taichenachev2006a,barber2006a}. Mastering the laser technology required to observe such transitions with a spectroscopic resolution at or below the Hertz level enabled the development of optical atomic clocks (see \cite{ye2008a,poli2013a} for reviews of this field, and \cite{porsev2004a,hong2005a,hoyt2005a,barber2006a,taichenachev2006a,poli2008a} for specific realizations using ytterbium atoms). Although optical atomic clocks are typically operated using dilute samples far from quantum degeneracy, many-body effects are nevertheless measurable because of the extreme spectroscopic sensitivity \cite{martin2013a}. 

A number of proposals have been made to use such ``clock" transitions, virtually free from spontaneous emission, to manipulate quantum-degenerate gases, with diverses purposes ranging from quantum information processing \cite{gorshkov2009a,daley2011a} to the realization of effective magnetic fields \cite{gerbier2010a} or strongly-interacting impurity models \cite{gorshkov2010a} in optical lattices. In this article, we report on high-resolution spectroscopy on Bose-Einstein condensates (BECs) of ytterbium atoms, a first step towards more complex experiments using the clock transition. We use here bosonic $^{174}$Yb and demonstrate how spectroscopic features of the ${^1}S_0 \rightarrow {^3}P_0$ transition, observed with sub-kHz resolution, can yield informations about the velocity distribution. Previously, spectroscopy of trapped Yb condensates on another ultranarrow $J=0\rightarrow J'=2$ transition has been demonstrated \cite{yamaguchi2010a,kato2012a,shibata2014a}. This transition is sensitive to magnetic fields. As a result, the  ${^1}S_0 \rightarrow {^3}P_0$ transition is expected to offer a better spectral resolution, at the expense of a much weaker transition strength. Spectroscopy of trapped Yb Fermi gases in three-dimensional optical lattices was reported in \cite{scazza2014a,capellini2014a}.

The article is organized as follows. In Section \ref{sec:expsetup}, we describe our experimental setup used to generate the laser light probing the clock transition, including a high-finesse optical cavity used for locking the laser frequency. In Section \ref{sec:iodine}, we show how we calibrate the absolute cavity frequency using an auxiliary spectroscopic reference based on molecular iodine \cite{hong2009a}. Section \ref{sec:tofspectro} describes spectroscopy on Yb BECs, where the condensate is first released from the trap to reduce the interaction energy. The resulting spectra are dominated by the velocity distribution of the expanding condensate, allowing us to measure spectra with widths of a few kHz. A single-shot technique to locate the laser frequency is presented in Section \ref{sec:Doppler}. Finally, we conclude in Section \ref{sec:conclusion}.

\section{Experimental setup}

\label{sec:expsetup}

\begin{figure}
\includegraphics[width=8.3cm]{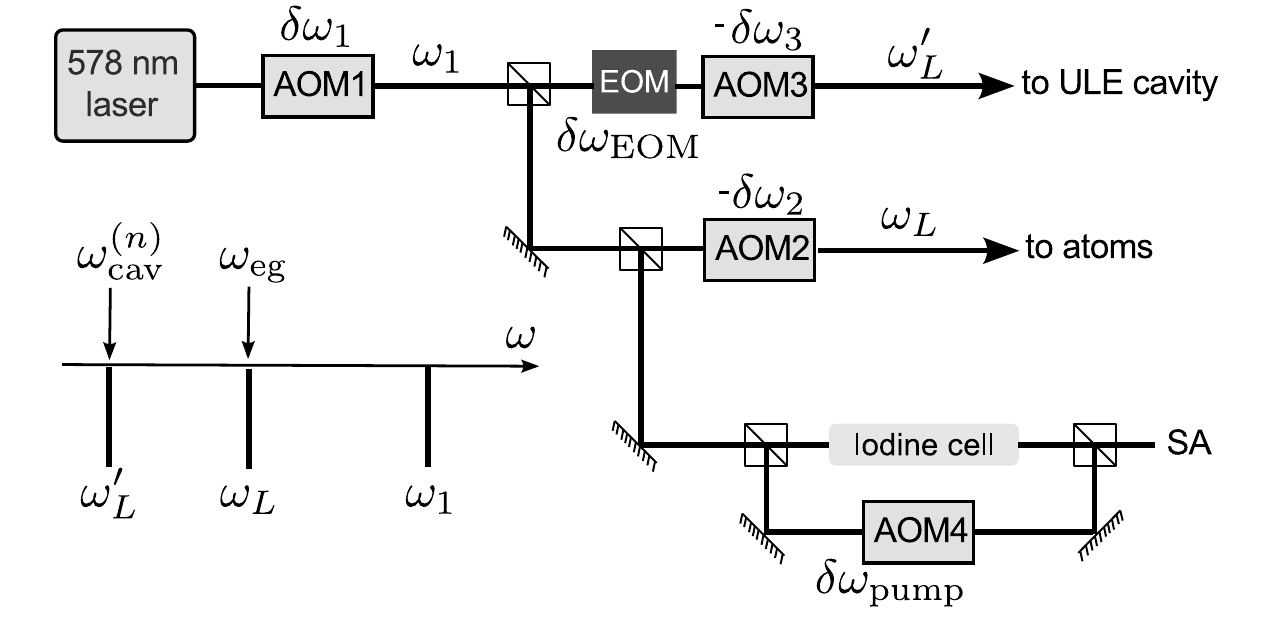}
\caption{Layout of the optical setup. A laser source at 578.4~nm is sent to three different paths, one leading to an ultrastable cavity to which the laser frequency is locked, one leading to the trapped atomic sample, and one used for iodine spectroscopy. The inset shows the spacing between the various frequencies involved, with $\omega_{\rm cav}^{(n)} $ the cavity mode to which the laser frequency is locked and $\omega_{eg}$ the clock transition frequency. AOM: acoustic-optical modulator. EOM: electro-optical modulator. ULE: ultra-low expansion glass. SA: saturated absorption.} \label{figure1}
\end{figure}	
	
	\subsection{Optical setup}\label{sec:optical}
	
	The main features of our experimental setup are shown in Figure~\ref{figure1}. Narrow-linewidth laser light with wavelength $\lambda_L=578.4$~nm is generated by sum-frequency mixing of two infrared lasers in a non-linear crystal \cite{oates2007a,hong2009a}. The infrared laser sources used in this work are a solid-state laser emitting around 200 mW near 1319~nm and an amplified fiber laser emitting around 500 mW near 1030~nm. The frequency of both lasers is controlled using a piezo-electric transducer (PZT) on the laser cavities. 
	The two lasers are coupled into an optical waveguide made from periodically-poled Lithium Niobate. The output of the waveguide is sent to a first acousto-optical modulator (AOM1), shifting the light frequency by a quantity $\delta\omega_1$. This AOM is used to control the laser frequency to ensure it remains resonant with the cavity, and to control its intensity. The laser is then split among three paths, a first one leading to an ultrastable optical cavity that we use as a stable frequency reference (see Section~\ref{sec:cavity}), a second one  leading to the atoms {\emph{via}} an optical fiber (see Section~\ref{sec:tofspectro}), and a third one to an iodine spectroscopy cell used to calibrate the absolute frequency of the cavity (see Section~\ref{sec:sas}). As detailed in Figure~\ref{figure1}, AOMs are used to control independently the frequencies of the light in each path. The cavity path includes an additional electro-optical modulator adding frequency sidebands near 1~GHz. Its purpose will be explained in the following. Single-mode, polarization maintaining optical fibers are used to transport the light from the optical bench where it is generated to the cavity and to the main experiment where ytterbium atoms are probed. We denote by $\omega_1$ the light frequency after passing through AOM1, by $\omega_L$ the frequency of the light used to probe the atoms and by $\omega_L'$ the frequency of the light sent to the cavity. 

	\subsection{Cavity design and characteristics}\label{sec:cavity}

We use an ultrastable Fabry-Perot optical cavity as frequency reference to stabilize the frequency of the laser used to perform spectroscopy on the clock transition. This type of cavities have been extensively studied in the past few years, in connection with the recent developments in optical atomic clocks \cite{nazarova2006a,ludlow2007a,alnis2008a,millo2009a,leibrandt2011a,leibrandt2011b}. The cavity we use is a commercial model from Advanced Thin Films (Boulder, Co), with a spherical body made from Ultra-low expansion (ULE) glass and two fused silica mirrors optically contacted to the ULE glass. The cavity is held at two points and placed inside a commercial housing (Stable Laser Systems, Co) consisting of the cavity holder surrounded by a gold-coated thermal shield, which are placed inside a vacuum chamber. This isolates the cavity from detrimental pressure and temperature changes. The temperature of the thermal shield is actively stabilized to a few milliKelvins using a Peltier element and a servo control. Using an additional EOM (not shown in Figure~\ref{figure1}), the laser frequency is locked to the cavity with the Pound-Drever-Hall technique \cite{hall1983a}. Frequency feedback is applied by changing the frequency $\delta\omega_1$ of AOM1, driven by a frequency synthesizer with fast modulation input and wide modulation range (Agilent Technologies model E4400B). An additional feedback path is used to correct for long term drifts by acting on the frequency-control input of the 1319~nm pump laser. 

The cavity resonances are labeled by a longitudinal mode index $n$, according to $\omega_{\rm cav}^{(n)} = \omega_{\rm cav}^{(0)} + n \cdot \Delta_{\rm FSR}$, where $\Delta_{\rm FSR}/2\pi= c/2L$ is the cavity free spectral range (FSR), $L=47.6(1)$~mm  is the cavity length and $c$ the speed of light in vacuum. The free spectral range separating two cavity resonances was calibrated using the EOM shown in Figure~\ref{figure1}, which adds sidebands spaced by $\sim1~$GHz to the laser spectrum. We performed a linear scan of the laser frequency and recorded the frequencies for which the first positive sideband (frequency $\omega_L'+\delta\omega_{\rm EOM}$) is resonant with cavity mode $n+1$ and the second negative sideband (frequency $\omega_L'-2\delta\omega_{\rm EOM}$) is resonant with cavity mode $n$. Taking the difference between the two measurements leads to the cavity FSR, $\Delta_{\rm FSR} = 2\pi \times 3~144~366(4)$~kHz (this measurement was done for a cavity temperature $T_{\rm cav}\approx 5.0~^\circ$C). 

The width of each cavity resonance is determined by the cavity finesse, dependent on the transmission of the cavity input coupler and other losses due to optical imperfections, diffraction, etc ... The finesse is conveniently obtained by measuring the cavity ring-down time by switching off the incoming laser and monitoring the decay of the power transmitted through the cavity over time. This measurement gives an exponential decay with $1/e$ decay time constant $\tau \approx 13~\mu$s. Relating this time constant to the cavity finesse $\mathcal{F}$ using the formula $ \tau^{-1} =\Delta_{\rm FSR}/\mathcal{F}$, we find $\mathcal{F} \approx 260~000$, which is consistent with the specifications of the manufacturer.

\section{Absolute frequency calibration of the clock laser using high-resolution spectroscopy on molecular iodine}\label{sec:iodine}

\begin{figure*}
\centering
\begin{tabular}{lr}
\includegraphics[scale=0.9]{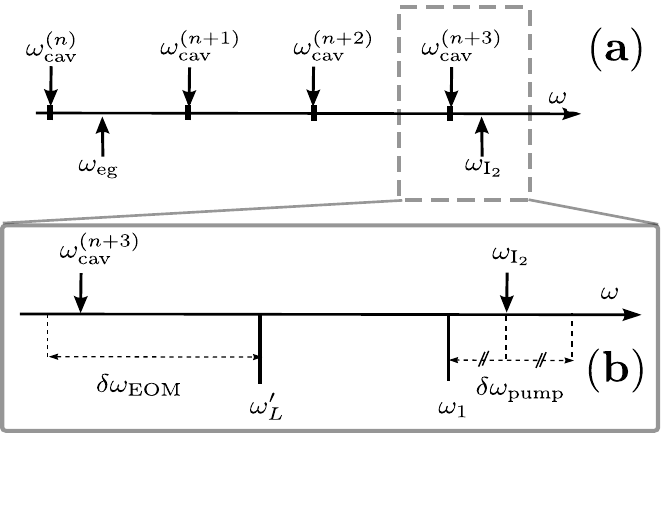}& \includegraphics[]{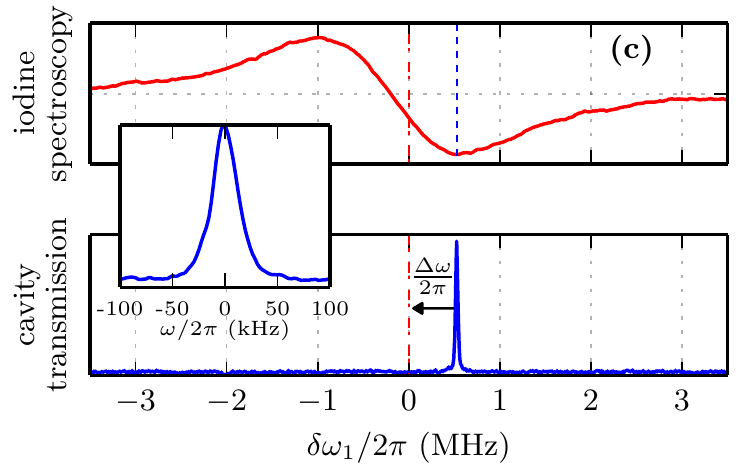}
\end{tabular}
\caption{ {\bf (a)}: Frequency ruler showing the relative positions of several cavity resonances label with mode index $n$ to $n+3$, the clock transition frequency labeled $\omega_{eg}$ and the iodine resonance labeled $\omega_{I_2}$. {\bf (b)}: Close-up near the iodine resonance, showing the relative positions of the different laser frequencies used for the spectroscopy when the iodine spectroscopy laser is resonant. {\bf (c)}: Experimental iodine absorption spectrum (top) and cavity transmission (bottom) as a function of laser frequency. The inset shows a zoom on the narrow cavity resonance.} 

\label{figure2}
\end{figure*}	
	
	\subsection{Motivation}\label{sec:mot}
	
	When dealing with ultra-narrow lines, a key experimental issue is to be able to locate the resonance. For $^{174}$Yb atoms and the ${^1}S_0 \rightarrow {^3}P_0$ transition (hereafter denoted "clock transition"), accurate values of the transition frequency have been measured by optical atomic clocks \cite{hong2005a,hoyt2005a,barber2006a,taichenachev2006a,poli2008a}. Measuring the laser frequency to the same precision is however difficult, as the absolute cavity frequency is {\emph {a priori}} unknown. A basic method would be to perform first a coarse measurement of the cavity frequency (which can be done with a typical precision on the order of hundred MHz using commercial wavemeters) and then systematically search for the atomic resonance in the interval corresponding to the frequency uncertainty. Needless to say, this is a cumbersome procedure, which would need to be done again each time the cavity resonance varies (intentionally or due to uncontrolled drifts). A more precise method to calibrate the cavity frequency is clearly desirable. In principle, this is achievable with modern technology combining frequency combs with stable microwave atomic clocks \cite{hollberg2005a}. 
	
	When this technology is not available, an alternative is to use an atomic or molecular line for frequency reference. Fortunately, in the case of the ${^1}S_0 \rightarrow {^3}P_0$ transition in $^{174}$Yb, a nearby resonance in molecular iodine has already been identified and characterized with an uncertainty of $\sim$2~kHz \cite{hong2009a}. This molecular resonance, located approximately $10~$GHz to the blue of the Ytterbium clock transition, corresponds to one particular hyperfine component of the so-called R(37)16-1 line. This $\Omega=0 \rightarrow \Omega'=0$ transition ($\Omega$ denotes the total molecular angular momentum) is essentially unaffected by magnetic fields and leads to narrow lines suitable to perform accurate frequency calibrations. In the following, we will denote this specific transition (corresponding to a frequency $\omega_{I_2}$) as ``iodine transition" for simplicity. We also label with ``$n$" the cavity resonance closest to the clock transition, and ``$n+3$" the one closest to the iodine transition located three FSRs away from the first one. The scheme to measure the absolute laser frequency is then clear : First, measure $\Delta \omega = \omega_{I_2}-\omega_{\rm cav}^{(n+3)}$, then deduce $\omega_{\rm cav}^{(n)}=\omega_{I_2}-\Delta\omega-3 \Delta_{\rm FSR}$ using the value $\omega_{I_2}$ measured in \cite{hong2005a} and the calibrated value for $\Delta_{\rm FSR}$, and finally deduce the frequency $\omega_L$ of the laser probing the atoms using the known AOM frequencies (see Figure ~\ref{figure2}a).

			\subsection{Saturated absorption spectroscopy}\label{sec:sas}
	
To observe the iodine resonance, we perform saturated absorption (SA) spectroscopy on a quartz cell filled with molecular iodine (lower path in Figure \ref{figure1}). We use a modulation transfer spectroscopy scheme : The probe beam at frequency $\omega_1$ travels unmodulated through the cell, while the counter-propagating pump beam first passes through an additional AOM, shifting its frequency by $+\delta \omega_{\rm pump}$ and modulating it for lock-in detection at the same time. In such a geometry, the saturated absorption resonance is reached when $\omega_{\rm I_2}=\omega_1+\delta \omega_{\rm pump}/2$. The particular line we investigate lies approximately $1.1~$GHz away to the red of the $n+3$ cavity resonance (see Figure~\ref{figure2}b). We use the EOM in the cavity path to bridge this frequency gap, tuning its modulation frequency so that the first negative sideband on the cavity path is near the $n+3$ cavity mode and the laser frequency is near the iodine resonance. In this way, we are able to measure both resonances within the same frequency scan. Typical SA and cavity transmission spectra are shown in Figure~\ref{figure2}c when scanning the laser frequency. By fitting the data (absorption and transmission) to Lorentzians, we extract the positions of the $n+3$ cavity resonance and of the iodine resonance, which differ by a quantity $\Delta \omega$, the outcome of the measurement. The standard deviation of 10 identical measurements performed with fixed parameters is below 10 kHz. 
		
	We have carefully looked for systematic effects that could affect this measurement. We have found no dependence on the power of the probe laser or applied magnetic field, but identified two effects that need to be accounted for in the frequency measurement. A first effect is instrumental. 
	The lock-in amplifier used to obtain the saturated absorption signal behaves as a first-order low pass filter with time constant $350~\mu$s. This results in a slight shift of the center of the dispersive lock-in signal shown in Figure~\ref{figure2}c from the iodine resonance. This is included in our analysis, where the expected Lorentzian line shape of the resonance is convoluted with the transfer function of the lock-in and fitted to the measured signal to extract the line position and width. 
	The second correction is due to collisions inside the iodine cell, which leads to line shift and broadening. We have carefully measured these effects and correct for them when evaluating the molecular resonance (see Appendix \ref{app:coll}). 
	
	Using this technique, we were able to quickly find the Yb resonance by interrogating a sample of ultracold atoms (see next Section). Repeating the measurements allows us to track changes of the cavity resonance frequencies over time, either because of intended changes in the parameters (\emph{e.g.} the temperature of the cavity), or because of uncontrolled environmental drifts. The error in the absolute frequency is larger than the precision of the measurement, and estimated to be a few tens of kHz. When comparing \emph{a posteriori} the laser frequency deduced from the measured iodine resonance with respect to the actual one found from the atomic spectra, we observe shifts of $\approx \pm 20~$kHz which are stable to a few kHz over extended periods of time, but can vary with modifications of the experimental setup. Possible explanations could be errors in determining the collisional correction, or cell contamination \cite{hong2009a}. According to the value given in \cite{barber2006a}, the differential light shift due to the probe laser itself on the clock transition is around $1~$kHz for our experimental parameters, and therefore cannot explain the observed difference between the observed resonance and the one predicted from the cavity calibration. 
	
	\subsection{Measurement of the zero-crossing of the cavity resonance with temperature}\label{sec:zerocross}

\begin{figure}
\includegraphics[width=8.3cm]{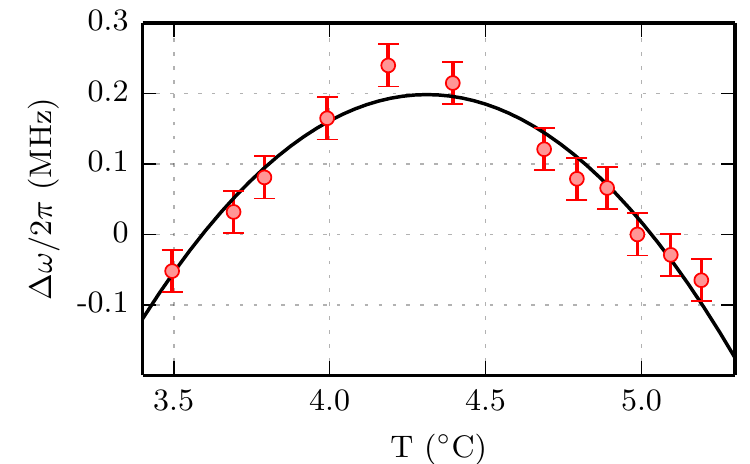}
\caption{Cavity resonance frequency (mode $n+3$) versus cavity temperature. The frequency is measured using iodine spectroscopy and reported in terms of the difference from an arbitrarily chosen reference point at $T_{\rm can}=5^\circ$C. The maximum corresponds to minimized sensitivity to environmental drifts.}
 \label{figure3}
\end{figure}

An important application of the spectroscopic calibration technique described above is the determination of the zero-crossing point of the cavity. This refers to the change of the resonance frequency with the cavity temperature, $d\omega_{\rm cav}^{(n+3)}/dT_{\rm cav}$, which quantifies the sensitivity to environmental drifts. We have performed a series of cavity calibrations using iodine spectroscopy after changing the temperature of the cavity and waiting for approximately one day for thermal equilibrium to settle (this was experimentally verified by monitoring the cavity frequency over time). The result is plotted in Figure~\ref{figure3} in terms of $\Delta \omega$, as defined in Section~\ref{sec:sas}. We are able to identify a clear maximum near $T_{\rm cav}\approx 4.3^\circ$C, where the linear thermal expansion coefficient vanishes. Working near this point is clearly desirable to improve the quality of the frequency reference provided by the cavity, as demonstrated in the next Section.
	
\section{Doppler spectroscopy of expanding Bose-Einstein condensates}
\label{sec:tofspectro}

\subsection{Optical spectrum of a trapped Bose-Einstein condensate}

We now turn to the main goal of this work, namely spectroscopic interrogation of bosonic $^{174}$Yb atoms on the clock transition. We work with a Bose-Einstein condensate created via forced evaporative cooling \cite{scholl2014a} in a crossed optical dipole trap. The trap is formed by two orthogonal beams, a first one at wavelength 1070 nm propagating along the horizontal $y$ axis and a second one at 532~nm propagating along $x$. This results in an approximately harmonic potential seen by the atoms, with frequencies $(\omega_x,\omega_y,\omega_z) \approx 2 \pi \times (130,260,290)~$s$^{-1}$. The condensate contains around $N \approx5.8 \times10^4$ atoms with an uncondensed fraction $< 20 \%$. The condensate is initially populated with atoms in the electronic ground state. Excitation to the metastable $^{3}$P$_0$ state is done using a laser beam with a Gaussian waist of $w\approx 30~\mu$m and an optical power $P\approx 1~$mW with vertical polarization. A vertical magnetic field of magnitude $B\approx 100~$G mixes a small amount of the nearby $^{3}$P$_1$ state to the $^{3}$P$_0$ state, thereby opening a transition channel on this otherwise ``doubly forbidden" transition \cite{taichenachev2006a, barber2006a}. 

We have performed spectroscopy of atomic samples held in the crossed dipole trap, and typically observed 10 kHz-broad spectra with an asymmetric line shape. Optical interrogation schemes of Bose-Einstein condensates were previously considered in the context of Bose-Einstein condensation of atomic hydrogen \cite{fried1998a,killian1998a} and Bragg spectroscopy \cite{stenger1999a,zambelli2000a} (see also \cite{yamaguchi2010a}). Two effects affecting the lineshape and linewidth of the resonance were identified, namely Doppler broadening due to residual motion in the trap and mean-field broadening due to different interactions strengths for atoms in different excited states. Additional factors in our case are a differential position-dependent light shift caused by the trap potential \cite{yamaguchi2010a}, and the collisional instability of the excited state. Inelastic collisions involving at least one excited atom do not conserve the principal quantum number. Such collisions impart both collisional partners with a kinetic energy much larger than the trap depth and result in net atom losses at a rate $\gamma_{ge}=\beta_{ge} n_g$ and $\gamma_{ee}=\beta_{ee} n_e$, with $n_g$ (respectively $n_e$) is the ground (resp. excited) state density. For $^{174}$Yb, the rate constants $\beta_{ge}$ and $\beta_{ee}$ are not known, nor is the $s-$wave coupling constant $g_{\rm ge}$ describing $g-e$ interactions (they have been measured for fermionic isotopes \cite{ludlow2011a}). We note that the determination of $g_{\rm ge}$ from the measured line shift suggested in \cite{oktel2002a} works only in absence of inelastic collisions and differential light shifts. Rough estimates suggest that the mean field, potential energy and inelastic losses contribute each a few kHz to the line broadening, complicating the interpretation of the spectra and lowering the achievable precision on the line center.

\subsection{Time-of-flight spectroscopy}
\label{sec:tof}

\begin{figure*}
\includegraphics[width=16.6cm]{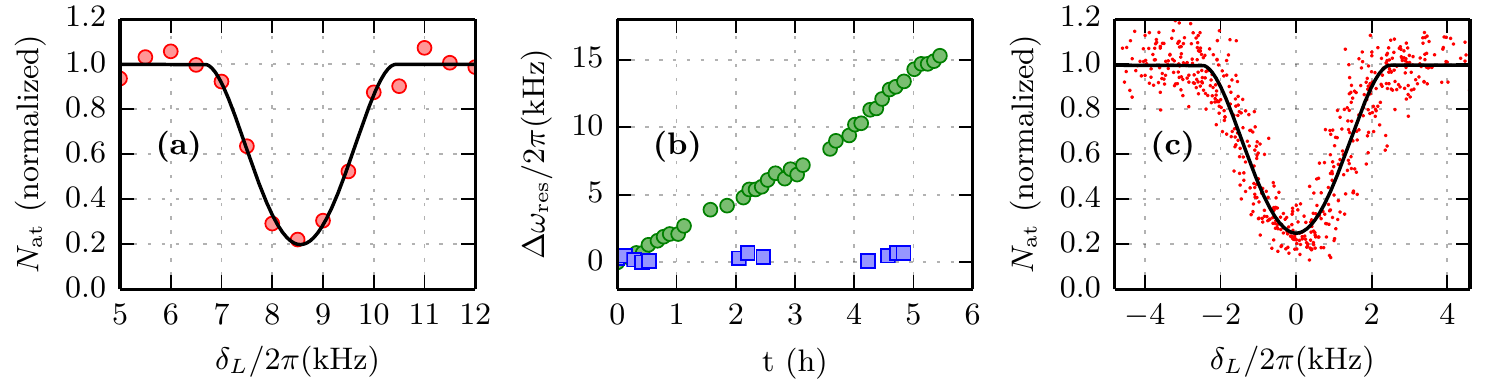}
\caption{{\bf (a)}: Optical spectrum of a $^{174}$Yb condensate probed in t.o.f. {\bf (b)}: Drift of the line center versus time, for a cavity far from the zero-crossing ($T_{\rm cav}\approx5.0^\circ$C, green circles) or close to it ($T_{\rm cav}\approx4.3^\circ$C, blue squares). {\bf (c)}: Drift-corrected ptical spectra for $T_{\rm cav}\approx5.0^\circ$C. The points shown correspond to all spectra in {\bf (b)} with $T_{\rm cav}\approx5.0^\circ$C recentered to correct for the drift.}
 \label{figure4}
\end{figure*}	

To minimize the role of interactions (both elastic and inelastic), we interrogate the atoms in free space after releasing them from the trap. We typically switch on the probe beam $200~\mu$s after releasing the atoms from the trap and leave it on for 3~ms. We then let the cloud expand freely in time of flight (t.o.f.) for $19~$ms and measure the remaining ground state population using standard absorption imaging. Atoms in the excited state are not detected : Successful excitation of the clock transition thus reduces the measured atom number. 

A condensate in a static harmonic potential $V_g({\bf r}) =\sum_{\alpha} \frac{1}{2}m\omega_\alpha^2 x_\alpha^2$ and in the Thomas-Fermi (TF) regime has a parabolic density profile, $n_g({\bf r}) = (\mu/g_{gg})\Upsilon \left[1- \sum_\alpha (x_\alpha/R_\alpha)^2 \right]$, with $\alpha=x,y,z$. Here we have defined a function $\Upsilon(x)=x$ if $x\geq 0$ and zero otherwise, the chemical potential $\mu$, the $s-$wave coupling constant $g_{gg}$ for ground state atoms and the TF radius along direction $\alpha$, $R_\alpha=\sqrt{2\mu/m\omega_\alpha^2}$. After the trap is suddenly switched off,  the time-dependent wave function obeys a scaling solution \cite{castin1996a,kagan1996a} where the density envelope
keeps the same parabolic form as in the trap with rescaled coordinates and TF radii, $R_\alpha(t)=R_\alpha b_\alpha(t)$. The scaling factors $b_\alpha$ obey the differential equations
$ \ddot{b}_\alpha +\omega^2_\alpha(t) b_\alpha=\omega_\alpha^2/(\mathcal{V}b_\alpha)$, with $\mathcal{V}=\Pi_\alpha b_\alpha$. The hydrodynamic velocity field is given by ${\bf v}=\sum_\alpha (\dot{b}_\alpha/b_\alpha) x_\alpha{\bf e}_\alpha $. 
This evolution can be interpreted as a conversion of the initial mean-field interaction energy (which tends to zero since the density drops as $1/\mathcal{V}$) into kinetic energy \cite{castin1996a,kagan1996a,pitaevskii2003a}. 

Assuming the expansion time is long enough (in practice after several oscillation periods), the mean-field pressure becomes negligible and the expansion becomes almost ballistic ($b_\alpha= \dot{b}_\alpha t$ with constant $ \dot{b}_\alpha$'s for $t\rightarrow \infty$). The resonance condition is then entirely determined by the Doppler effect,
\begin{align}
\delta_L=\omega_L- \omega_0= k_L v_x,
\end{align}
where $\omega_0 =\omega_{eg} +\omega_r$ is the sum of the free space Bohr frequency $\omega_{\rm eg}$ and of the recoil correction $\omega_r=\hbar k_L ^2/2m$. Here we chose a probe wave vector ${\bf k}_L=k_L {\bf e}_x$ and noted $k_L$ the laser wave vector. The Doppler-broadened optical spectrum is given by the velocity distribution (integrated along $v_y,v_z$) evaluated at the resonant velocity $v_x^\ast =\delta_L/k_L$ \cite{zambelli2000a},
\begin{align} \label{eq:A}
A(\delta_L) &= \frac{15 N}{16\Delta} \Upsilon\left[  1- \left(\frac{\delta_L}{ \Delta}\right)^2 \right]^2,
\end{align}
where the spectrum width is 
\begin{align} \label{eq:D}
\Delta =k_L R_x \dot{b}_x.
\end{align}  

In a first series of measurements, the cavity temperature was $T_{\rm cav}\approx 5.0~^\circ$C, far from the zero-crossing point. Individual spectra obtained with a single frequency scan (a few minutes), as shown in Figure~\ref{figure4}a, were fitted with Eq.~(\ref{eq:A}) with the resonance frequency, amplitude and width as free parameters. The resonance frequency $\omega_{\rm res}$ shows a substantial drift over time, presumably associated with the cavity (see Figure~\ref{figure4}b). The time dependence is well approximated by a piecewise linear function of time with slope $\sim 3~$kHz/hour. The impact on the spectra can be suppressed by correcting each spectrum for the measured center drift. This produces an averaged, drift-free spectrum shown in Figure~\ref{figure4}c. The typical measured linewith is $\Delta/2\pi \approx 2.5~$kHz. This compares well to the calculated width based in Eq.~\ref{eq:D}, $\Delta_\infty \approx 3.1~$kHz. The difference between the measurement and the asymptotic value is small, but significant. It could be explained by an underestimation of the atom number or by a deviation from a Doppler-dominated spectrum. A more complete theory should account for both the Doppler and interaction broadening in a time-dependent fashion. While the former dominates near the end of the probe pulse, the latter is not negligible in the beginning. We have not attempted such a detailed modeling, as our main interest is to monitor the central frequency and its evolutions. 

We have repeated these measurements after tuning the cavity to the zero-crossing temperature $T_{\rm cav}\approx 4.3~^\circ$C. In Figure~\ref{figure4}b, the effect of tuning the cavity to its zero-crossing point is clearly shown. The frequency drift of the laser (or equivalently of the cavity) is no longer observable. The resonance frequency shows a residual dispersion around the same value with a standard deviation of $0.2~$kHz. In this configuration, correcting for residual long-term changes of the resonance is no longer necessary for the current spectral resolution.

\subsection{Doppler spectroscopy of a condensate expanding in a waveguide}\label{sec:Doppler}

\begin{figure}
\includegraphics[width=8.3cm]{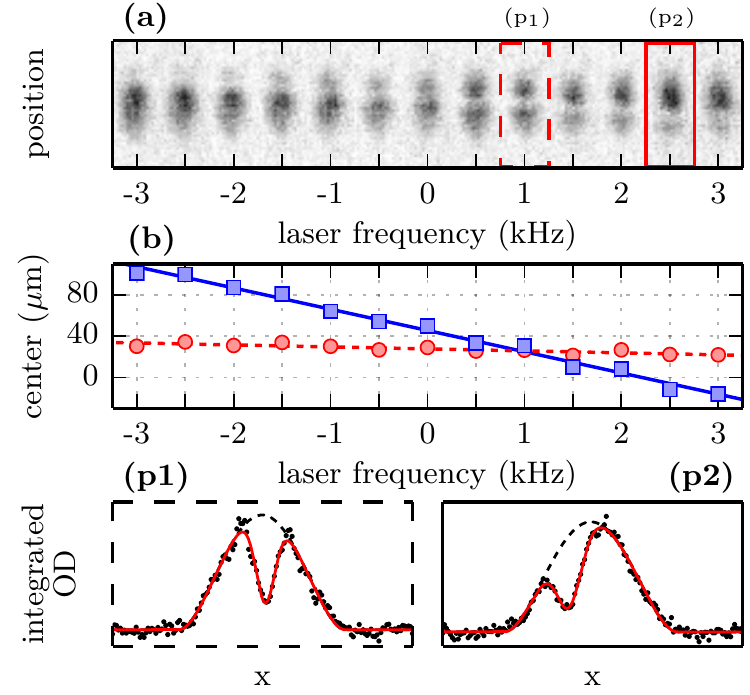}
\caption{Doppler velocimetry of an expanding BEC. {\bf (a)}: Experimental pictures versus probe laser frequency. {\bf (b)}: Position of the resonant slice (blue squares) and of the cloud center (red circles) versus frequency of the probe laser. The slice position is linear with the laser frequency, as expected when probing the velocity distribution (see text).  The lower panels {\bf (p1,p2)} show two profiles (integrated along the horizontal direction) together with the fit used to extract the cloud and ``hole" centers and widths.}
 \label{figure5}
\end{figure}

We have performed a second series of experiments where the ballistic nature of the condensate expansion is ensured by releasing it in a ``waveguide" instead of free space. This is done by switching the 1070~nm dipole trap off. The atoms then undergo a quasi-one dimensional expansion along $x$ inside the waveguide formed by the remaining dipole trap. This expansion is also described by a scaling solution, with $b_x \gg b_y,b_z$. After $10~$ms of expansion, the atoms are released in free space and we repeat the same probing sequence as before. Differently from the experiments described in the previous Section, the expansion along $x$ (the direction of propagation of the probe laser) becomes ballistic after a few ms only, well before we apply the probe pulse. These experiments were performed for $T_{\rm cav}\approx5.0^\circ$C.

\begin{figure}
\includegraphics[width=8.3cm]{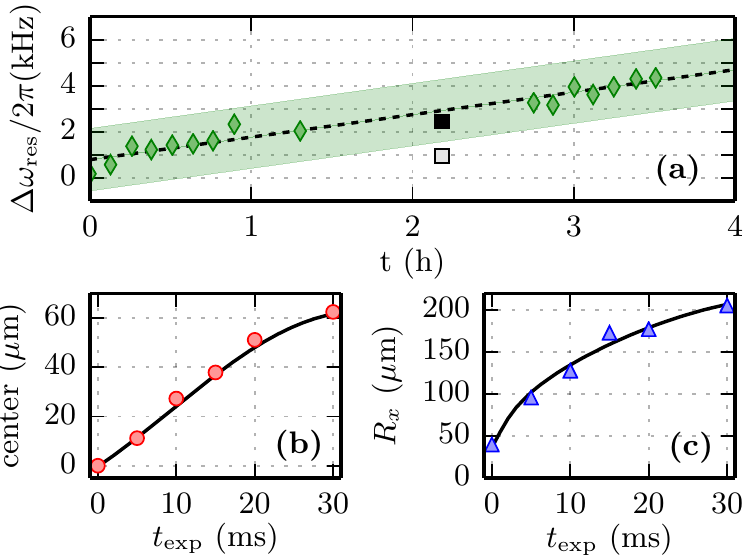}
\caption{ {\bf (a)}: Measured change of the resonance frequency $\omega_{\rm res}$ from t.o.f. spectroscopy (green diamonds) and from the slice positions before (gray square) and after (black square) correcting for the measured acceleration of the cloud. The light green shaded area shows corresponds to the full width at half-maximum of the t.of. spectrum. {\bf (b)}: Position of the cloud center (red circles) versus expansion time. {\bf (c)}: Width of the cloud center (blue triangles) versus expansion time. The solid lines in {\bf (b,c)} are fits to the data using the scaling model, leaving the waveguide trap frequencies as free parameters.}
 \label{figure6}
\end{figure}

The absorption images in Figure~\ref{figure5}a  show visually the position of the missing atoms as the laser frequency is scanned. Time of flight maps initial velocities to final positions, so that the ``slice" of missing atoms corresponds directly to the resonant velocity class $v_x^\ast$. We have fitted the profiles of the cloud (as shown in Figure~\ref{figure5}b) to a TF profile multiplied by an heuristic ``hole" function to account for the missing atoms. From this fit, we can extract the position $x^\ast$ of the hole and $x_0$ of the cloud center, which are plotted in Figure~\ref{figure5}b. One would have expected that the two should coincide near the center of the spectrum, which is not the case. This effect is explained by a center of mass motion (c.o.m.) of the cloud along $x$, presumably due to a misalignment of the trap foci (see Figure~\ref{figure6}b). The resonance condition for a moving cloud is given by
\begin{align} 
\label{eq:xres}
\delta_L = k_L \left( \frac{ \dot{b}_x}{b_x} (x^\ast-x_0)+\dot{x}_0\right),
\end{align}
with $\dot{x}_0$ the c.o.m. velocity. The hole position depends linearly on the laser frequency, with a measured slope $1/(2\pi)\times 20.6~\mu$m/kHz. We performed a fit to the data in Figure~\ref{figure6}c using the scaling model, leaving the waveguide trap frequencies as free parameters. From this fit, we deduce $b_x\approx 18$, $\dot{b}_x\approx 569$~s$^{-1}$. From Eq.~(\ref{eq:xres}), we find a slope $d x^\ast/d\omega_L\approx 1/(2\pi)\times 18.3~\mu$m/kHz that compares well with the measured one. 

When the hole and c.o.m. positions coincide, the resonance is Doppler shifted by the c.o.m. velocity, that can be measured and corrected for. Figure~\ref{figure6}c shows that when the c.o.m. correction is applied, the resonance frequency deduced from the position of the resonant slice agrees within $1~$kHz with the one determined in Section~\ref{sec:tofspectro}, given the linear drift in the resonance frequency observed during the data acquisition (see Section~\ref{sec:tofspectro}). 

For future experiments, our results suggest the possibility to measure the absolute laser frequency in one shot by locating the slice position in a given image, which greatly facilitates tracking the possible drifts and changes of the cavity for quantum gases and quantum information experiments. The resolution of the current setup can be improved greatly with small changes in the experimental setup. For instance, a frequency resolution of a few Hz should be possible by probing a cloud released from cigar-shaped trap with frequencies $(\omega_x,\omega_y,\omega_z) \approx 2 \pi \times (20,400,400)~$s$^{-1}$ and $N=3\cdot 10^4$, for which we find from the scaling theory a frequency width $\Delta /2 \pi \approx 65~$Hz. The main practical limit will be the free fall of the cloud, which limits the transit time through the probe beam but can in principle be eliminated with a suitable optical potential levitating the atoms against gravity. 

\section{Conclusion}
\label{sec:conclusion}

In conclusion, we have performed high-resolution spectroscopy on a Bose-Einstein condensate of ytterbium atoms using an ultranarrow optical transition. The transition is probed by a laser locked on a high-finesse optical cavity. We have shown that the cavity frequency could be calibrated within a few tens of kHz using a nearby iodine absorption line, which greatly facilitates locating the resonance on a day-to-day basis. Experiments on expanding BECs were presented, and interpreted in terms of Doppler-sensitive spectroscopy probing the velocity distribution. We proposed a quantitative description based on the scaling solution \cite{castin1996a,kagan1996a} that describes well the experimental observations. Finally, we showed a technique where the laser frequency can be determined in a single image of an expanding BEC. 

Hydrodynamic expansion of a BEC is well-established \cite{pitaevskii2003a}, and our experiment allowed us to verify the expected features. This suggests that high-resolution spectroscopy is a promising tool to explore many-body systems. The current resolution of a few kHz is limited by mean-field effects, which can be reduced or modified by working in optical lattices or in elongated traps. Our current experimental system has a resolution of a few hundred Hertz, limited by Doppler phase noise added by the optical fibers transporting light to the cavity and to the atoms. Such noise can be suppressed by an appropriate servo-loop \cite{ma1994a}. In combination with sufficient isolation of the cavity from vibrations, a resolution on the order of $10~$Hz should be feasible, which is comparable to the resolution achievable with microwave spectroscopy \cite{campbell2006a}. Optical spectroscopy as demonstrated here has the additional feature that it is sensitive to the atomic momentum, and therefore allows one to probe different and physically relevant quantities, such as the spectral function in an interacting system (bosonic or fermionic) \cite{dao2009a}.

\begin{acknowledgments}
We acknowledge many stimulating discussions with members of the BEC and Fermi gases groups at LKB, of the Frequency metrology group at SYRTE (Observatoire de Paris), in particular P. Lemonde, R. Le Targat and Y. Le Coq. and with M. Notcutt (Stable Laser Systems). We thank Thomas Rigaldo for experimental assistance with the 578 nm laser. We acknowledge financial support from the ERC under grant 258521 (MANYBO) and from the city of Paris (Emergences program). M. Scholl is supported by a fellowship from UPMC. 

\end{acknowledgments}
\appendix
\section{Collisional effects in Iodine spectroscopy}
\label{app:coll}
As well-known, near room temperature, absorption spectra in molecular iodine are substantially modified by collisions \cite{fletcher1995a}. Collisions cause a line broadening $\delta \omega_{\rm B}$ and a line shift $\delta \omega_{\rm S}$ with respect to the free space resonance, which depend on the iodine partial pressure $P_{I_2}$ and on the temperature $T$ as $\delta\omega_{\rm S/B} =2\pi  \alpha_{S/B} x$ with $x=P_{I_2}T^{-7/10}$. 
The line shift and line broadening are expected to be proportional to the collision rate $\gamma_{\rm coll} =  n_{I_2} \sigma \overline{v}$, where $n_{I_2} $ is the iodine density inside the cell, where $\sigma$ is the collisional cross-section and where $\overline{v}$ is the mean thermal velocity. The linear pressure dependence is thus natural, but the temperature dependence can seem odd at first glance. It originates from the dependence of the collisional cross-section on the velocity in the quasi-classical limit \cite{landaumq}, which for a Van der Waals interaction is  $ \sigma \propto \overline{v}^{-2/5}$. One then has $\gamma_{\rm coll} \propto n_{I_2} \overline{v}^{3/5} \propto x$. 

A cold finger at the bottom of the cell is maintained at a temperature of $\approx 5.0~^\circ$C by a Peltier element, which allows us to control the partial pressure of iodine inside the cell. Using the known vapor pressure curve for iodine \cite{gillespie1932a}, we found that the measured line shift and line width were indeed linear with $x=P_{I_2}T^{-7/10}$, where $T_{\rm cell}\approx20~^\circ$C is the temperature of the cell and where $P_{I_2}$ is evaluated at the temperature of the cold finger. We find $\alpha_B \approx 4.5(4)~$MHz.K$^{7/10}$.Pa$^{-1}$ and $\alpha_S \approx -400(60)~$kHz.K$^{7/10}$.Pa$^{-1}$, in agreement with earlier measurements \cite{fletcher1995a}. 

We correct the measured line center from the collisional line shift as follows. The measured line is assumed to be determined by the convolution of two Lorentzians of a background width $\Gamma_0/2\pi \approx 2~$MHz and of the collisional broadening,  $\Gamma=\Gamma_0 + \alpha_B x$. Note that $\Gamma_0$ is not limited by the natural line width of a few hundred kHz. For a given spectrum giving the line center $\omega_{\rm SA}$ and the line width $\Gamma_{\rm SA}$, we extract $x=(\Gamma_{\rm SA}-\Gamma_0)/\alpha_B$ and correct the iodine resonance frequency according to $\omega_{I_2}=\omega_{\rm SA}-\alpha_S x$.


%
%
\end{document}